\documentclass[aps,twocolumn,prb,reprint,amsmath,amssymb,showpacs,superscriptaddress]{revtex4-1}

\usepackage{graphicx}
\usepackage{dcolumn}
\usepackage{bm}
\usepackage{amsmath,amsfonts,amsthm}
\usepackage{color}
\usepackage{mathrsfs}
\usepackage{float}
\usepackage{ulem}
\usepackage{txfonts}
\usepackage{wasysym}
\usepackage{cancel}

\makeatletter

\newcommand{\Rmnum}[1]{\expandafter\@slowromancap\romannumeral #1@}
\makeatother
\begin{document}
\title{Signatures of triplet superconductivity in $\nu=2$-chiral Andreev states} 

\author{Liliana Arrachea}
\affiliation{Escuela de Ciencia y Tecnolog\'{\i}a  and ICIFI, Universidad Nacional de San Mart\'{\i}n, Av 25 de Mayo y Francia, 1650 Buenos Aires, Argentina}
\author{Alfredo Levy Yeyati}
\affiliation{
Departamento de F{\'i}sica Te{\'o}rica de la Materia Condensada, Condensed Matter Physics Center (IFIMAC)
and Instituto Nicol{\'a}s Cabrera, 
Universidad Aut{\'o}noma de Madrid, 28049 Madrid, Spain}
\author{C. A. Balseiro}
\affiliation{
Centro At\'{o}mico Bariloche and Instituto Balseiro, 8400 Bariloche, Argentina}

\begin{abstract}
We study the behavior of the conductance and the current-noise in three-terminal configurations of edge modes of a quantum Hall system
in the $\nu=2$ filling factor with normal and s-wave superconducting contacts. We discuss the impact of spin-orbit coupling in the quantum Hall system
and the possibility of effectively inducing triplet pairing in the egde states. We show that the presence of these correlations imprints very clear signatures in both the non-linear conductance and noise in these type of devices.
\end{abstract}

\date{\today}
\maketitle

\section{ Introduction.} 
The  coexistence of the superconductivity with the quantum Hall regime and the peculiar nature of the chiral Andreev states
that develop in the  edge states when contacted to superconductors  motivated several  works for some time now
\cite{ma1993josephson,eroms95,hoppe2000andreev,giazotto2005andreev,akhmerov2007detection,stone2011josephson,van2011spin}.
The search for realization of topological superconductivity  with p-wave  pairing 
\cite{alicea2012new,beenakker2013search,qi2011topological} provided an extra boost to the study of such hybrid structures. In fact, one of the proposed platforms to realize this phase in two-dimensional structures relies on the hybridization of a quantum anomalous Hall system with an s-wave superconductor \cite{qi2010chiral}. This strategy is akin to contacting the edge states of the 
quantum-Hall state to s-wave superconductors \cite{mong2014universal,clarke2014exotic}. These ideas heightened the interest
in studying the exotic properties of these systems and resulted in a notable upsurge  in both experimental \cite{wan2015induced,amet2016supercurrent,park2017propagation,lee2017inducing,draelos2018investigation,
seredinski2019quantum,zhi2019coexistence,guiducci2019full,guiducci2019toward,zhao2020interference,hatefipour2022induced,vignaud2023evidence}
and theoretical \cite{gamayun2017two,gavensky2020majorana,
gavensky2021imaging,manesco2022mechanisms,kurilovich2022criticality,kurilovich2023disorder,tang2022vortex,david2023geometrical,cuozzo20234} endeavors.

Several of these experiments focus on graphene \cite{amet2016supercurrent,park2017propagation,lee2017inducing,draelos2018investigation,seredinski2019quantum,zhao2020interference,vignaud2023evidence,zhao2023loss}, since this material has the advantage 
 of requiring low magnetic fields, which favors the coexistence of the quantum Hall regime with the superconductivity. 
However, experiments in hybrid structures with superconductors where the two dimensional electron system (2DES) is realized in other materials  such as  InAs and InSb have been also reported \cite{eroms95,batov2007,guiducci2019toward,guiducci2019full,zhi2019coexistence,
hatefipour2022induced}. In such a scenario, spin-orbit coupling (SOC) is expected to play a relevant role.
The theoretical description of SOC in a 2DES under the quantum Hall regime has been discussed in Refs.  \cite{Pala2005rashba,reynoso2004theory}. The combination 
with superconductors has been discussed 
in the spin-polarized $\nu=1$-filling factor \cite{van2011spin,michelsen2023supercurrent}. There, it was shown that the Rashba SOC in the interface between the 2DES and the superconductor in combination with the magnetic field leads to an effective
 p-wave type of pairing in the chiral edge mode. 

\begin{figure}[htb]
\centering
\includegraphics[width=\linewidth]{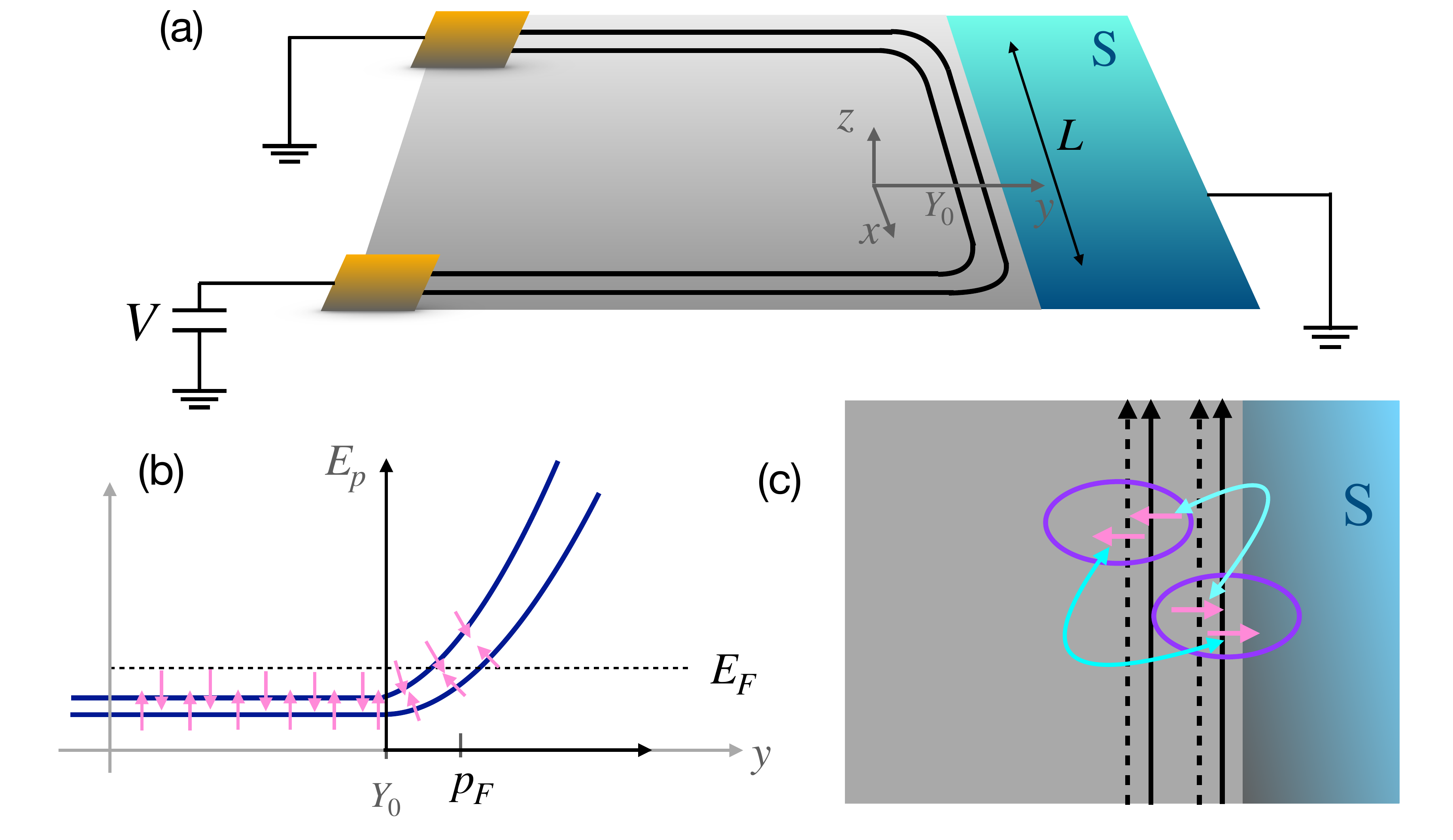}
     \caption{(a) Sketch of the setup. The 2DES in the quantum Hall state in the $\nu=2$ filling factor is contacted within a length $L$ with a grounded superconductor terminal.
     The edge states are also connected through normal Ohmic contacts to source and drain terminals at potentials $V$ and zero, respectively. (b) Profile of Landau levels and edge states with a dispersion relation consistent with a spin-orbit coupling effectively inducing p-wave-type intra edge  pairing on the edge states. (c) Pairing processes induced by proximity on the edge states. Without SOC, only
     s-wave pairing inter-edge exists (see light-blue arrows). The effect of the SOC is to induce additional triplet pairing (see violet elipses). }
     \label{fig0}
 \end{figure}

In the present work we  analyze  configurations where a s-wave superconductor is proximitized to a 2DES in the quantum Hall regime with filling factor $\nu=2$.
This is the lowest $\nu$ hosting chiral Andreev states induced by proximity with s-wave superconductors for which experimental results have been reported. 
We show that the interplay between the magnetic field, the SOC and the superconductivity induces superconducting pairing with both s-wave and p-wave-type components in the edge states even when the SOC exclusively affects  the 2DES. This is a realistic scenario for compounds based on In \cite{hatefipour2022induced}, where SOC is expected to be strong in contrast to graphene, where it is thought to be weak.
Importantly, we demonstrate that 
non-linearities in the
dispersion relation of the edge states lead to the development of p-wave superconductivity. We consider a setup with three terminals 
 --two normal Ohmmic contacts and the superconductor-- with a voltage bias applied at one of the normal contacts as sketched in Fig. \ref{fig0}. We calculate the conductance within and beyond the linear response as well as the current noise at the drain normal terminal.
 We show that the behavior of these quantities provides crucial insight into the nature
of the pairing induced at the edge states. Specifically, the presence of p-wave paring reveals itself through non-linear response in both the conductance and the noise. This phenomenon becomes a distinctive hallmark of the elusive p-wave superconductivity.


\section {Model} 
Our first goal is the derivation of an effective Hamiltonian for the edge states of the 2DES under
 the $\nu=2$-quantum Hall regime  for the configuration 
 sketched in Fig. \ref{fig0} (a) with the s-wave superconductor contacted in a region of length $L$.  The 2DES is in the $(x,y)$ plane under the effect of a magnetic field in the $z$-direction, which induces a Zeeman field in this direction,
 in addition to the orbital magnetism. The 2DES 
is also subject to a SOC of the Rashba type, induced by the electric field $\vec{E}= E_0 \vec{z}$ which is expected for this geometry. Such  interaction
is described in terms of the following Hamiltonian 
\begin{equation}\label{soc}
    H_{\rm SOC}= -  \frac{\mu_B}{c^2}  \left( \vec{\rm v}_p \times \vec{E} \right) \cdot \vec{S},
\end{equation}
where $\mu_B$ is the Bohr magneton, while $\vec{S}$, $\vec{\rm v}_p={\rm v}_p \vec{x}$ and $m$ are the spin, the velocity  and the mass of the electron, respectively.
 
\begin{figure*}[t]
\centering
\includegraphics[width=\linewidth,height=0.3\textheight]{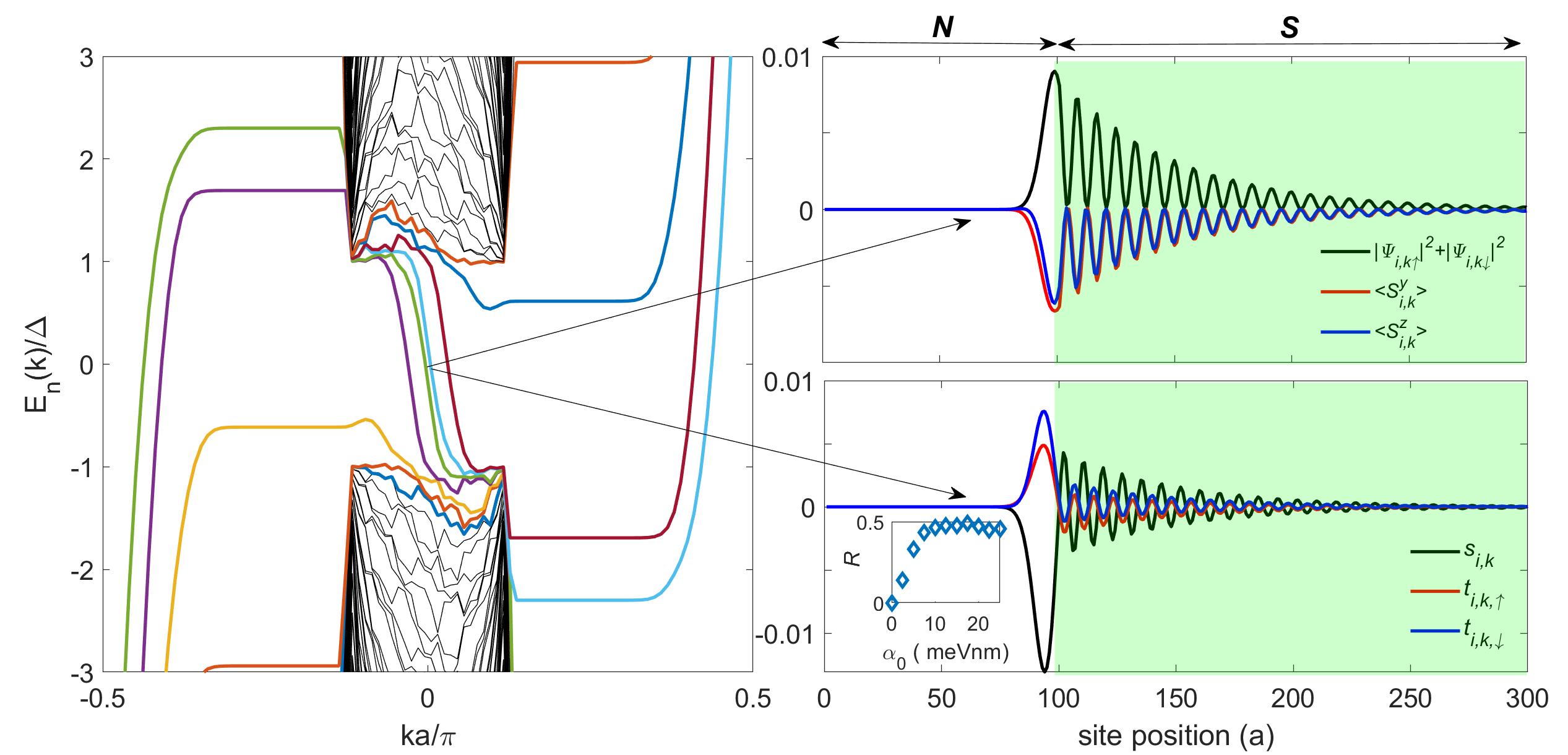}
     \caption{Left: Spectrum calculated with exact diagonalization of a discretized BdG Hamiltonian on a square lattice (with lattice parameter $a$) with a magnetic flux and SOC within the normal region and local s-wave pairing in the superconducting region (see text).  Periodic boundary conditions are considered along $x$. Right: Probability density of the chiral Andreev states
     indicated with the arrow in the left panel, along  the direction $y$ together with the weight of the polarization components $\langle S_y\rangle$ and $\langle S_z\rangle$ (top) and  the functions $t_{i,k,\sigma}$ and $s_{i,k}$ characterizing the pairing amplitudes in the triplet and singlet channels (bottom). Inset: ratio $R$ between of the space-averaged triplet and singlet amplitudes as a function of the strength of the SOC, as defined in the text.} For details on the parameters see Appendix \ref{numerical}.
     \label{fig:fig1}
 \end{figure*}

 The evolution from the spectrum of bulk
 Landau levels to edge modes has been the subject of many studies \cite{wei1998edge,zhang2013local,pascher2014imaging,panos2014current,patlatiuk2018evolution}. It is usual to describe the  edge states in terms of
 a linear dispersion relation. It is however known that in 
some phenomena, such as thermalization of edge states, the deviation from linear dispersion is found to 
play a crucial role \cite{karzigPhysRevLett.105.226407,karzig2012relaxation,bard2018long}. Here, we focus on the impact of 
such non-linear effects on the SOC and the consequences on the induced pairing. We  consider edge modes with  the  dispersion relation sketched in Fig. \ref{fig0} (b)
and
 we substitute Eq. (\ref{soc}) by the expansion with respect to the Fermi momentum $p_F$
\begin{equation}\label{vel}
    {\rm v}_p= \frac{dE_p}{dp} \simeq \rm{v} + \delta {\rm v} (p-p_F),
\end{equation}

  Under the effect of the external Zeeman field and the SOC the 
Hamiltonian matrix for the edge modes can be expressed in the basis 
$\left(c_{p,\uparrow}, c_{p,\downarrow}\right)$ with the spin quantization axis along $z$. It 
reads,
\begin{equation}\label{model2}
H_{0}(p) =  {\rm v} \sigma^0 (p-p_F)  - B_{Z} \sigma^z - B_{\lambda} \sigma^y
- {\rm v}_\lambda \sigma^y \left(p-p_F\right),
\end{equation}
where the first term represents the kinetic energy assuming the usual linear dispersion with velocity ${\rm v}$, corresponding to the velocity
at the Fermi momentum.  The second term represents the Zeeman field. The SOC is described in terms of an effective magnetic field $B_{\lambda}= \mu_B {\rm v} E_0/( mc^2) $,
which corresponds to substituting the constant term of the velocity in Eq. (\ref{soc}) and a spin-dependent correction to the kinetic term corresponding to
the correction $\delta {\rm v}$. The latter is encoded in the parameter
$ {\rm v}_\lambda= \mu_B \delta {\rm v} E_0/( mc^2)$. 
Reported calculations for the Landau levels in the presence of SOC and a confinement potential generating the 
edge states are fully consistent with this picture \cite{Pala2005rashba,reynoso2004theory}.  Hamiltonian \eqref{model2} can be diagonalized 
and the corresponding
eigenstates  define the scattering states injected from the source and exiting at the end of the superconducting contact towards the drain (see Fig. \ref{fig0}).

We now consider the effect of the pairing correlations induced on the edge modes by the proximity to the superconductor. 
The Hamiltonian describing the local s-wave pairing  reads
\begin{equation}
H_\Delta = \int_0^L dx \left(\psi^{\dagger}_\uparrow(x) \Delta_{s}\psi^{\dagger}_\downarrow(x)
+\rm{H.c.}\right),
\end{equation}
where the field operators describe the electrons with spin $\sigma=\uparrow,\downarrow$ in the position $x$ along the edge. It is  important to stress that the projection of this pairing in the basis that diagonalizes Eq. (\ref{model2}) has singlet-type inter edge and triplet-type intra edge components (see Appendix \ref{pairing}).
The corresponding amplitudes read, respectively,
\begin{eqnarray}\label{pairings}
\Delta_{0} & \simeq &
B_{Z} \frac{\Delta_s }{B_0},\;\;\;\;\;\;\;
\Delta_{T,p} 
\simeq -{\rm v}_\Delta p.
\end{eqnarray}
We have introduced the definition ${\rm v}_\Delta= \Delta_s {\rm v}_\lambda/B_0$, being
$B_0=\sqrt{B_Z^2+B_\lambda^2}$.


The resulting Bogoliubov de Gennes (BdG) Hamiltonian   expressed in the basis that diagonalizes Eq. (\ref{model2}) reads
\begin{eqnarray}\label{model2-1}
H_{\rm BdG}(p) &=& \tau^0 \otimes   \left[{\rm v}   p \sigma^0  -B_0 \sigma^z \right] -  \varepsilon_0\tau^z \otimes \sigma^0 \\
& & + \left\{{\rm v}_{\Delta}(x),  p\right\} \tau^x \otimes \sigma^y + \Delta_0(x) \tau^x, \nonumber
\end{eqnarray}
where $\sigma^j, \; \tau^j, j=0,\ldots,3$ are $2 \times 2$ unit matrices ($j=0$) and the three Pauli matrices ($j=1,2,3$) acting, respectively, on the spin (with the quantization axis along $\vec{B}_0$) and particle-hole degrees of freedom. 
Here,
 the pairing functions ${\rm v}_{\Delta}(x)$ and $\Delta_0(x)$ are non-vanishing in the finite region $0\leq x\leq L$.
We also introduced the parameter $\varepsilon_0$, which takes into account that the Fermi level of the 2DES can be slightly shifted away from
${\rm v}p_F$
without changing the filling factor by recourse to a gate voltage. Finally, $ \left\{{\rm v}_{\Delta}, p\right\} $ denotes the anticommutator, which accounts for the spacial dependence of ${\rm v}_{\Delta}$ in terms of a hermitian operator. This Hamiltonian defines the effective  model for energies lower than $\Delta_s$
to describe the edge states of the 2DES with SOC in proximity with the superconductor.

 In compounds like those studied in Ref. \cite{hatefipour2022induced} the SOC acts on the full 2DES. The starting point in our
derivation of Eq. (\ref{model2-1})  was an effective model for the edge modes in the presence of SOC, including
terms beyond the usual linear dispersion relation of these modes. In what follows, we
 benchmark the validity of our conclusions   against results of  numerical calculations based on a 2D lattice Hamiltonian for the full structure, accounting for 
 the effect of a magnetic field, the
 SOC and the proximity with the s-wave superconductor.  This model is obtained by discretizing the BdG equations describing the 2DES contacted with the superconductor (see Appendix \ref{numerical} for details). 
In these calculations periodic boundary conditions in the $x$-direction (parallel to the boundary between the two systems) are imposed. 
The 2DES is defined in the region denoted by "N" in Fig. \ref{fig:fig1}. In this region, 
a Peierls phase accounts for the magnetic flux and a Zeeman field is also considered.  The remaining sites define the superconductor (region "S" in the Fig.), where the Hamiltonian has a local s-wave  pairing. The effect of the SOC is described by a Rashba Hamiltonian in the 2DES with a modulating function $1/2-\tanh\left[(y-y_b)/\xi_{\lambda}\right]/2$, being $y_b$ the position of the
boundary between the 2DES and the superconductor and $\xi_{\lambda}$ a characteristic length of a few lattice sites that describes a smooth transition decay of the SOC into the superconductor. The chemical potential is fixed to have the 2DES in the $\nu=2$ filling factor, corresponding to the state where the two Zeeman levels of the lowest Landau level are filled. The spectrum of the Bogoliubov-de Gennes Hamiltonian is shown in the left panel of  Fig. \ref{fig:fig1} as a function of the wave vector $k$ defined along the $x$ direction. We can clearly identify the two pairs of particle-hole chiral  Andreev states. As mentioned before, due to the combination of the Zeeman field and the effective SOC field the spin of these states  has components $\langle S_z \rangle$ and $\langle S_y \rangle$. The upper right panel of Fig.\ref{fig:fig1} shows the behavior of these components along the $y$ direction and across the interface
for one of the edge states. For the other chiral Andreev state we observe a similar behavior (see. Fig. 1 of Appendix. \ref{numerical}). We notice that both states overlap in space with similar weights and opposite signs of $\langle S_z \rangle$ and $\langle S_y \rangle$.  This behavior is in agreement with the
description of the effective model of Eq. (\ref{model2-1}).
In the bottom-right panel of Fig. \ref{fig:fig1} we analyze
the singlet and triplet pairing components of these states. To this end we define the function
 $s_{i,k}=\langle c_{i,k,\uparrow} c_{i,-k,\downarrow}-c_{i,k,\downarrow} c_{i,-k,\uparrow}\rangle$ as a measure of the 
singlet pairing in the state $k$ at the lattice site with coordinate $i$ along the $y$-direction. Similarly,  the functions $t_{i,k,\uparrow}=\langle c_{i,k,\uparrow} c_{i,-k,\uparrow}\rangle$ and $ t_{i,k,\downarrow}=\langle c_{i,k,\downarrow} c_{i,-k,\downarrow}\rangle$ are signatures of triplet pairing. The behavior of these quantities for the  state at zero energy 
(indicated with arrows in the Fig.) is in full agreement with the effective Hamiltonian. Furthermore, we can see in the inset that the ratio $R=\sum_i \left(|t_{i,k,\uparrow}|^2+|t_{i,k,\downarrow}|^2\right)/\sum_i|s_{i,k}|^2$ goes to zero as the intensity of the SOC vanishes, in agreement with Eq. (\ref{pairings}).

\section{Transport properties}
Having verified the validity of the effective Hamiltonian for the edge states defined in Eq. (\ref{model2-1}), we now focus on the analysis of the transport properties generated by a bias voltage $V$ at the source reservoir. We rely on this model to calculate the conductance associated to the current entering the  drain reservoir as well as the associated noise. 

The current can be expressed in terms of the transfer matrix $M(E)$ relating the outgoing states
with respect to the superconductor (for $x>L$) with the incoming ones (for $x<0$).  It reads
(see Appendix \ref{transport})
\begin{equation}
J=\frac{e }{2h}  \sum_{\alpha=1}^4 \int dE {\cal M}^{\alpha,\alpha}(E,E)f_{\alpha}(E),
\end{equation}
being
${\cal M}(E,E^{\prime}) = M^{\dagger}(E) \; \tau^z\otimes \sigma^0 \; M(E^{\prime})$.
Here $\alpha$ labels the four components of the spinor associated to the incoming electrons. Hence,
$f_{\alpha}(E)= 1/\left(1+e^{(E -\mu_{\alpha})/k_BT} \right)$
 is the Fermi function corresponding to the the temperature $T$ and the bias voltage for the particle and hole components. Respectively,
 $\mu_{1}= \mu_2= eV$ and $\mu_{3}= \mu_4= -eV$.

Expanding the Fermi functions we get the expressions for the linear and non-linear components of the conductance
from $J=\sum_{n=0}^\infty G^{(2n+1)} V^{2n+1}$. We introduce the definition of the transmission function
\begin{equation} \label{trans}
{\cal T}(E)= \frac{1}{2}\mbox{Tr}\left[\tau^z\otimes \sigma^0 {\cal M}(E,E)\right],
\end{equation}
in terms of which the different orders of the conductance at $T=0$ read
\begin{equation}\label{cond}
G^{(2n+1)}=\frac{e^{2}}{h}\frac{1}{(2n+1)!}\frac{d^{(2n)} {\cal T}(E)}{dE^{(2n)}}|_0.
\end{equation}
We notice that only the odd powers in $V$ are non-vanishing.

Following a similar procedure we calculate the noise corresponding to the current-current correlation 
(details are presented in SM, Appendix.\ref{apnoise}). It can be expressed as follows
\begin{eqnarray}\label{S-fin}
S (eV)
&= &  \frac{e^2}{4 h^2} \sum_{\alpha,\overline{\alpha}} \int dE  {\cal M}^{\alpha \overline{\alpha}}(E, E) {\cal M}^{\overline{\alpha} \alpha} (E,E) 
F_{\alpha}(E),
\end{eqnarray}
where we have introduced the definition $F_{\alpha}(E) =f_{\alpha}(E) \left[ 1- f_{\overline{\alpha}}(E) \right]$. 
The non-vanishing combinations in the sum are $\alpha=1,2; \;\; \overline{\alpha}=3,4$.



The transfer matrix for the Hamiltonian of Eq. (\ref{model2-1}) is  (see Appendix \ref{det-her})
\begin{eqnarray}\label{transfer}
M(E) &=& 
\exp \left\{ \frac{i L E {\rm v}}{\hbar\tilde{\rm v}^2}  \right\}
 \exp \left\{  \frac{iL}{\hbar \tilde{\rm v}}   \left[ \varepsilon_0 \tau^z\otimes \sigma^0 -B_{0} \tau^0\otimes  \sigma^z       \right.\right.\nonumber\\
 & & \left.\left. \;\;+ \;
 {\Delta}_{0}\;
\tau^x\otimes \sigma^0
-\frac{ {\rm v}_{\Delta} E}{\tilde{\rm v}} \tau^x \otimes \sigma^y
 \right] \right\}, 
\end{eqnarray}
with $\tilde{\rm v}=\sqrt{{\rm v}^2-{\rm v}_{\Delta}^2}$, assuming ${\rm v} > {\rm v}_{\Delta}$.

\section{Results}
 It is useful to analyze the limiting cases of pure singlet-type inter-edge
 pairing, corresponding to $\Delta_0 \neq 0,\; {\rm v}_{\Delta}=0$ and pure triplet pairing, corresponding to $\Delta_0 = 0,\; {\rm v}_{\Delta}\neq 0$. Although the latter limit cannot be achieved in the model defined from
 Eq. (\ref{pairings})  we analyze  it here as a reference. 
 In these cases we can derive the following analytical expressions for the transmission function. For the {\it pure singlet} case we have 
 \begin{equation} \label{taus}
    {\cal T}_s=
      \frac{2}{r_s^2}\left[\varepsilon_0^2 + \Delta_0^2\cos\left(\frac{2L}{{\rm v} \hbar }r_s\right)\right],\;\;\;\;\;\; {\rm v}_{\Delta}=0,\; \forall B_0, 
 \end{equation}
 $r_s=\sqrt{\varepsilon_0^2+\Delta_0^2}$. 
  We clearly see that the transmission function is independent of $E$, which implies a purely linear conductance 
 \begin{equation}\label{gs}
 G_s^{(1)}={\cal T}_s G_0, \;\;\;\;\;\;\;\;\;\; G_s^{(2n+1)}=0, \;n\neq0.
 \end{equation}
 This is
 expected to display oscillations as a function of $\varepsilon_0$ for fixed values of the parameters $L, \Delta_0$. Such oscillations will become sizable for 
 $L > \xi_0 $, being $\xi_0= (\hbar {\rm v})/\Delta_0$, the effective superconducting length on the edge. Unlike the usual transmission function for 
 normal systems, ${\cal T}_s$ displays changes in the sign as a function of the gate voltage represented by $\varepsilon_0$. This striking feature is a consequence of the exotic nature of these Andreev states,
 which consist of an interference of particles and holes propagating chirally. This peculiar behavior has been reported in experimental studies \cite{zhao2020interference,hatefipour2022induced}.

 Instead, for the {\it pure} triplet-type intra-edge case the transmission  function depends on $E$ and reads
  \begin{equation} \label{taut}
   {\cal T}_{t}(E)  = \sum_{\sigma=\uparrow, \downarrow}\frac{1}{r_{t,\sigma}^2}\left[\varepsilon_\sigma^2 + \left(\frac{{\rm v}_{\Delta} E}{\tilde{\rm v}} \right)^2  \cos\left(\frac{2Lr_{t,\sigma}}{\tilde{\rm v} \hbar }\right)\right],\;\; {\Delta}_0=0, 
 \end{equation}
 with $\varepsilon_{\uparrow,\downarrow}=\varepsilon_0 \pm B_0$ and
 $\;r_{t,\sigma}=\sqrt{\varepsilon_\sigma^2+\left(\frac{{\rm v}_{\Delta}}{\tilde{\rm v}} E\right)^2}$.
 Remarkably, ${\cal T}_t (0)=2$, which implies that the linear conductance 
 is always equal to the ideal conductance quantum per channel for any value of  $\varepsilon_0, \; B_0$. The other remarkable feature is the fact that the non-linear conductance is non-vanishing. Explicitly, the linear conductance and the lowest non-linear component read
 \begin{equation}\label{gt}
 G_t^{(1)}=2G_0, \;\;\;\;\;\;\; G_t^{(3)}=-\frac{8 G_0}{3}  \left(\frac{{\rm v}_{\Delta}}{\tilde{\rm v} \varepsilon_\sigma}\right)^2 \sin^2(\frac{L | \varepsilon_\sigma|}{\tilde{\rm v} \hbar}).
 \end{equation}
  The analysis of the behavior of the transmission function ${\cal T}(E)$ in cases with both singlet and triplet type of pairing is presented in Appendix \ref{det-her}.
 
The current noise exhibits also a very different behavior in these two limits. While it is a linear as a function of $V$ for pure singlet pairing,
 it is fully non-linear for the pure triplet case. 

\begin{figure}[htb]
\centering
\includegraphics[width=\linewidth]{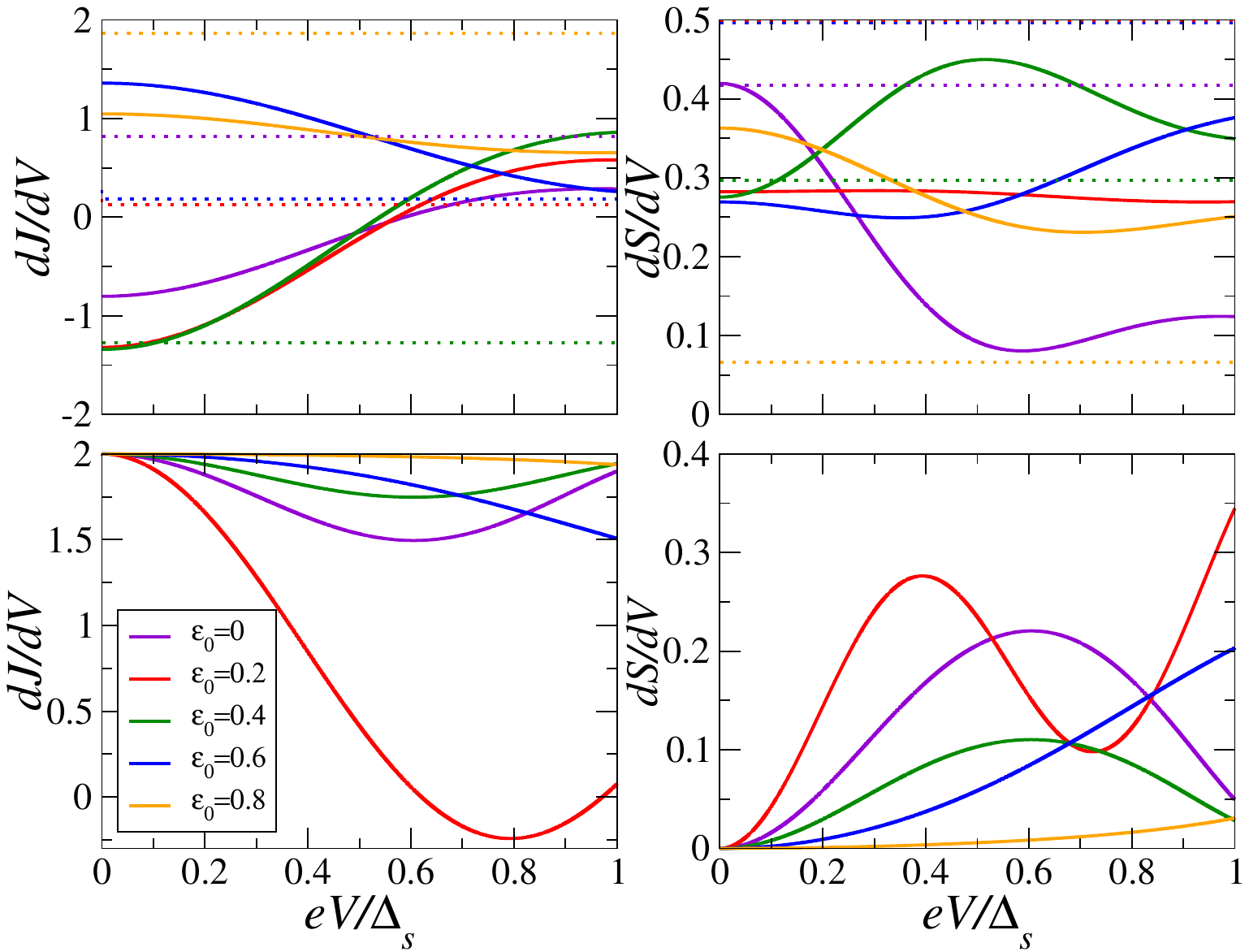}
     \caption{Conductance and noise as a function of the bias voltage $V$ at temperature $T=0$, for a superconducting contact of width $L=10 \xi_0$ and $B_Z=0.2 \Delta_0$.
     $\Delta_0=\Delta_s=1$ and ${\rm v}_{\Delta}/{\rm v}=0.2$. Dashed lines correspond to ${\rm v}_{\Delta}/{\rm v}=0$. Bottom panels correspond to
     $\Delta_0=0$. Different plots correspond to different values of the gate voltage represented by $\varepsilon_0$. All the energies are expressed in units of $\Delta_s$.}
     \label{fig:fig3}
 \end{figure}

 The  non-linear conductance $dJ/dV$ as well as the noise $dS/dV$ are shown, respectively, in the left/right panels of Fig. \ref{fig:fig3} for temperature $T=0$. The upper panels of the figure correspond to a system with both singlet and triplet components in the pairing. As a reference, we show in dashed lines  the corresponding (constant) values for pure singlet pairing  defined in Eq. (\ref{gs}). The limit of pure triplet pairing is shown in the bottom panels, where we see that the conductance
 approaches the limit $G_t^{(1)}$ defined in Eq. (\ref{gt}), while  $dS/dV$ vanishes as $V\rightarrow 0$. The behavior of these quantities in the case of both types of pairing has features of the two limiting cases. In fact, the conductance
 becomes flat as $V\rightarrow 0 $ and tends to a value different from $2G_0$. Furthermore, it may achieve positive as well 
 as negative values as $\varepsilon_0$ changes, as is the case of pure singlet pairing. Albeit, the values at $V=0$ are different from the ones for $v_\Delta=0$ shown in dashed lines. For large $V$, the non-linear response clearly emerge. 
 The behavior of $dS/dV$ is also different from zero for $V\rightarrow 0$, as in the case of pure singlet pairing. As $V$ increases, the non-linear features are also clear in the behavior of the noise.




 \section{Conclusions and discussion} 
 We have shown that  in a two-dimensional electron system under the $\nu=2$-quantum Hall regime in proximity to a s-wave superconductor, s-wave-type inter-edge as well as p-wave-type intra-edge pairing is induced in the chiral edge states as a consequence of the spin-orbit coupling.

 We have calculated the transport properties  and we have identified the hallmark of the two types of pairing components in the behavior of the conductance and of the current noise. 
 The most remarkable feature introduced by the triplet pairing is the development of a non-linear response in both the conductance and the noise, which could be easily identified in experimental studies. This component is originated in the intra-edge
 pairing induced into the chiral edge modes under the influence of the spin-orbit coupling. We have shown that a fundamental ingredient in this scenario is to take into account the
 non-linear dispersion relation of these states. This intra-edge triplet component coexists with the singlet inter-edge component. The latter generates a linear 
 response in the conductance with a
 peculiar positive or negative sign, which could be tuned by means of a gate voltage.

So far, only signatures of such singlet inter-edge type of pairing have been detected in experiments carried out  in  $\nu=2$-quantum Hall systems in proximity to superconductors \cite{amet2016supercurrent,hatefipour2022induced}. 
In fact, it is important to notice that only the behavior of the linear conductance has been reported in
these works. Our results constitute a motivation for the analysis of non-linear features in 
In-based devices and other systems, where spin-orbit coupling is expected to play a role. 
We expect that such non-linear features should be robust and amenable to be observed in experimental samples hosting spin-orbit coupling.

\section{Acknowledgements}
We thank stimulating conversations with Felix von Oppen, Stefan Heun, Matteo Carrega and Alexander Mirlin.
Support from CONICET as well as FonCyT, Argentina, through grants  PICT-2018-04536 and PICT 2020-A-03661 (LA), 
PICT 2016-0791, PICT 2018-1509, PICT 2019- 0371 (CB), and grant PIP-CONICET 11220150100506 (CB) are acknowledged. ALY
acknowledge support from Spanish AEI through grant PID2020- 117671GB-I00.

\appendix

\section{Induced pairing in edge states with SOC and a Zeeman field} \label{pairing}
Eq. (3) can be diagonalized by the transformation
\begin{equation}\label{new-basis}
\left(\begin{array}c
c_{p,\uparrow} \\
c_{p,\downarrow}
\end{array}
\right)=
\left(
\begin{array}{cc}
u_{p} & -v_{p} \\
v_{p} & u_p
\end{array} \right)
\left(\begin{array}c
\tilde{c}_{p,+} \\
\tilde{c}_{p,-}
\end{array}\right)
\end{equation}
with
\begin{eqnarray}
    u_p&=& \frac{1}{\sqrt{2}} \sqrt{1+\frac{B_{Z}}{r_p}},\nonumber \\
     v_p&=& \frac{i}{\sqrt{2}} \mbox{sgn}[\lambda_p]  \sqrt{1-\frac{B_{Z}}{r_p}},   
\end{eqnarray}
being $r_p=\sqrt{\lambda_p^2+B^2_{Z}}$.

In the transformed basis the Hamiltonian  reads
\begin{equation}\label{h0diag}
    H_0 =\sum_{s=\pm} E^0_{p,s} \tilde{c}^\dagger_{p,s} \tilde{c}_{p,s},
\end{equation}
with
\begin{eqnarray}
    E^0_{p,s}&=&{\rm v} \left(p-p_F\right) +s \sqrt{[{\rm v}_\lambda (p-p_F)+B_{\lambda}]^2+B_Z^2} \nonumber \\
   & &  \simeq {\rm v}  \left(p-p_F\right) +  s B_0.
\end{eqnarray}
We have introduced the definition $B_0=\sqrt{B_Z^2+B_\lambda^2}$ and in the last step we assumed  $B_0 \gg {\rm v}_\lambda (p-p_F)$.

We now consider the effect of the pairing term induced by the proximity to the superconductor described by the Hamiltonian 
\begin{equation}
H_\Delta = \Delta_0 \sum_p \left(c^{\dagger}_{p,\uparrow} c^{\dagger}_{-p,\downarrow}-c^{\dagger}_{p,\downarrow} c^{\dagger}_{-p,\uparrow}
+\rm{H.c.}\right).
\end{equation}
Substituting the change of basis it 
can be written as follows,
\begin{eqnarray}
    & & H_\Delta =  \sum_{p,s}\left[\Delta_{S,p}  \left(s \tilde{c}^{\dagger}_{p,s} \tilde{c}^{\dagger}_{-p,-s}+ \rm{H.c.}\right)+
    \Delta_{T,p}  \left(\tilde{c}^{\dagger}_{p,s} \tilde{c}^{\dagger}_{-p,s}+ \rm{H.c.}\right)
    \right],\nonumber
\end{eqnarray}
which describes pairing in the singlet inter-edge and triplet intra-edge channels.
The corresponding amplitudes read, respectively,
\begin{eqnarray}\label{pairings1}
\Delta_{S,p} &=& \Delta_s \left(u_p^2-v_p^2\right) =  \Delta_0 \frac{B_Z }{r_p}  \nonumber\\
\Delta_{T,p} &=& -2 \Delta_s u_p v_p = - \Delta_0 \frac{ \lambda_p}{r_p},
\end{eqnarray}
which reduce to Eq. (5) for dominant $B_0$.
\vspace{1cm}

\section{Numerical simulations}\label{numerical}
To analyze the properties of the chiral Andreev states at the interface between a spin-orbit coupled 2DES in the quantum Hall regime and a proximitixed superconducting region we discretize the corresponding BdG equations in a square lattice (with lattice parameter $a$), which leads to the following model Hamiltonian
\begin{widetext}
\begin{eqnarray}
    H_{2D} &=& \sum_{i=1,k}^{N_{\rm tot}} \Psi_{i,k}^{\dagger}\left[\left(2t\cos(ka+\tau_z\phi_i)-\mu_i - 4t\right)\tau_z\sigma_0 - 2\alpha\sin(ka+\tau_z\phi_i)\sigma_y\tau_0 
     - \Delta_i\tau_x\sigma_z + V_{i,Z} \sigma_z\tau_0 \right] \Psi_{i,k}   \nonumber \\
     && + \sum_{i=1,k}^{N_{\rm tot}-1} \Psi_{i,k}^{\dagger} \left(t\tau_z\sigma_0 + i\alpha\sigma_x\tau_0\right) \Psi_{i+1,k}  + \mbox{h.c.}
\end{eqnarray}
\end{widetext}
where $\Psi_{i,k}=(c_{i,k,\uparrow} \; c_{i,k,\downarrow} \; c^{\dagger}_{i,-k,\downarrow} \; c^{\dagger}_{i,-k,\uparrow})^T$; and
$t=-\hbar^2/(2m^* a^2)$ is a nearest neighbor spin-conserving hopping determined by the discretization parameter $a$ and the effective mass $m^*$, $\alpha_i$ is a spin-flipping hopping amplitude determined by the Rashba spin-orbit coupling in each region, $\mu_i$ and $\Delta_i$ are the local chemical and pairing potentials respectively, $V_{i,Z}$ is the Zeeman field and $\phi_i$ is the Peierls phase determined by the applied field. In this model we have assumed periodic boundary conditions on the $x$ direction, so that the momentum $\hbar k$ in this direction is conserved. Within this model the first $N_n$ sites correspond to the normal region, where $\Delta_i=0$, $V_{i,Z}=V_{Z}$ and $\mu_i=\mu_n$, while the rest $N_{\rm tot} - N_n$ sites correspond to the superconducting region, where $\Delta_i=\Delta$, $V_{i,Z}=0$ and $\mu_i=\mu_s$. 

We assume that the spin-orbit coupling varies as
\[ \alpha_i = \frac{\alpha_0}{4a}\left[1 - \tanh\left(\frac{(i-N_n)a}{\xi_{\lambda}}\right)\right] \;\]
where $\xi_{\lambda}$ is a characteristic length of a few lattice sites describing a smooth decay of the SOC inside the superconducting region. 

On the other hand, the magnetic field is assumed to be finite only in the normal region so that
\[ \phi_i = \left( \begin{array}{c} \frac{\phi}{N_n}(i-N_n) \;\; \mbox{if} \;\; i \le N_n \\
                                    0   \;\; \mbox{if} \;\; i > N_n \end{array} \right. \]
where $\phi$ is the total flux in the normal region in units of the flux quantum.

\begin{figure*}[t]
\centering
\includegraphics[width=\linewidth,height=0.3\textheight]{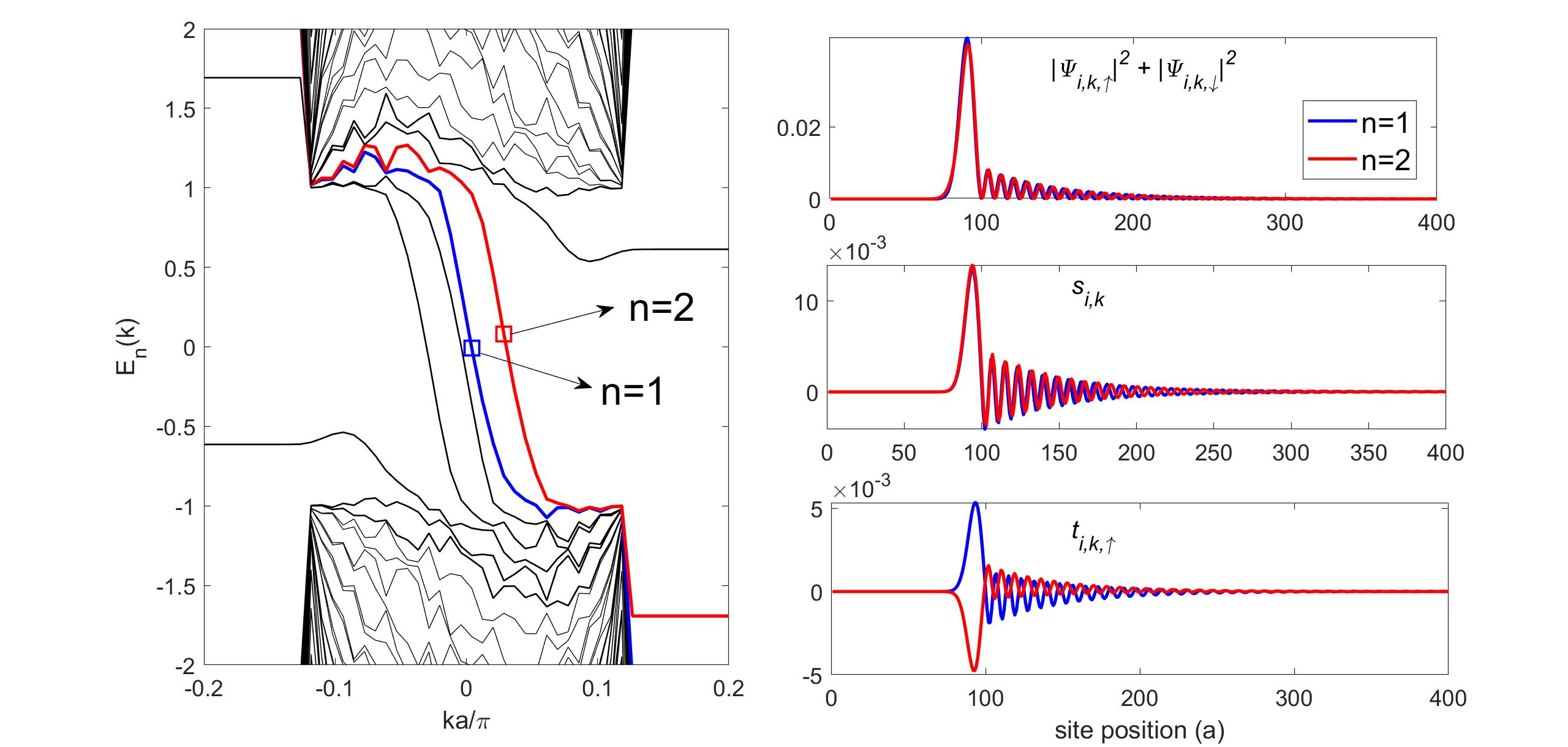}
     \caption{Left: Spectrum calculated with exact diagonalization of a discretized BdG Hamiltonian on a square lattice (with lattice parameter $a$) with a magnetic flux and SOC within the normal region and local s-wave pairing in the superconducting region (see definitions in the main text).  Periodic boundary conditions are considered along $x$. Right: Probability density of the chiral Andreev states
     indicated with the arrows in the left panel (top) and amplitudes of the pairing potential in the singlet (middle) and triplet channels (bottom). } 
     \label{fig:fig1s}
 \end{figure*}

For the calculations in Fig. 1  we have used: $N_{\rm tot} = 400$, $N_n=100$, $a = 2 {\rm nm}$, $m^*=0.08$, $\mu_n = 3.3 {\rm meV}$, $\mu_s = 18.8 {\rm meV}$, $\alpha_0=25 {\rm meVnm}$, $\Delta = 1 {\rm meV}$, $V_Z = 0.2 {\rm meV}$ and $\phi=1.75\pi/4$. 
These are the same as in Fig. 1 of the main text.

\section{Details on the calculation of the conductance}\label{transport}
 The current entering the drain reservoir is 
\begin{equation}\label{j}
J= \frac{e {\rm v} }{2}  \sum_{\alpha=1}^4 \langle \left[\Psi_{o}(t)^{\dagger}\tau^z \otimes \sigma^0 \Psi_{o}(t) \right]_{\alpha,\alpha}\rangle,
\end{equation}
where the field operator $\Psi_{o}(t)=\left(\psi_{o,\uparrow}(t),\psi_{o,\downarrow}(t), \psi^{\dagger}_{o,\downarrow}(t), -\psi^{\dagger}_{o,\uparrow}(t) \right)^T $ is defined for the edge states, in the region $x>L$, between the end of the scattering region and the drain normal contact.
We introduce the representation 
\begin{equation}\label{oi}
\Psi_{o}(t)=  \int \frac{dE}{\sqrt{h {\rm v}}}  e^{-\frac{i}{h} E t}  \Psi_{o}(E)=  \int \frac{dE}{\sqrt{h {\rm v}}} e^{-\frac{i}{h} E t} M(E) \Psi_{i}(E),
\end{equation}
being $M(E)$ the transfer matrix relating incoming and outgoing particles. We define the matrix
\begin{equation}
{\cal M}(E,E^{\prime}) = M^{\dagger}(E) \tau^z M(E^{\prime}).
\end{equation}

Substituting Eq. (\ref{oi}) in Eq. (\ref{j}) we get
\begin{equation}\label{jj}
J= \frac{e }{2h}  \sum_{\alpha,\alpha^{\prime}=1}^4 \int dE dE^{\prime} {\cal M}^{\alpha,\alpha^{\prime}}(E,E^{\prime}) \langle \left[\Psi^{\alpha}_{i}(E)\right]^{\dagger}  \Psi^{\alpha^{\prime}}_{i}(E^{\prime}) \rangle,
\end{equation}

In these expressions $\Psi^{\alpha}_{i/o}$ denotes the component $\alpha$ of the spinor $\Psi_{i/o}$ defined previously.
We now take into account that
\begin{eqnarray}
\langle \left[\Psi^{\alpha}_{i}(E)\right]^{\dagger}  \Psi^{\alpha^{\prime}}_{i}(E^{\prime}) \rangle &=& f_{V}(E) \delta_{\alpha, \alpha^{\prime}} \delta(E-E^{\prime}), \;\;\;
\alpha=1,2,\nonumber \\
\langle \left[\Psi^{\alpha}_{i}(E)\right]^{\dagger}  \Psi^{\alpha^{\prime}}_{i}(E^{\prime}) \rangle &=& f_{-V}(E) \delta_{\alpha, \alpha^{\prime}} \delta(E-E^{\prime}),\;\;\;
\alpha=3,4.
\end{eqnarray}

Hence, after some algebra we get
\begin{equation}
J=\frac{e }{2h}  \sum_{\alpha=1}^4 \int dE {\cal M}^{\alpha,\alpha}(E,E)f_{\alpha}(E).
\end{equation}
$f_{\alpha}(E)= 1/\left(1+e^{(E -\mu_{\alpha})/k_BT} \right)$
 is the Fermi function corresponding to the the temperature $T$ and the bias voltage $\mu_{\alpha}=\pm eV$ with $+, (-)$ for $\alpha=1,2,(3,4)$, respectively.

In order to fulfill conservation of the current, it should be satisfied
\begin{equation}
\sum_{\alpha=1}^2{\cal M}^{\alpha,\alpha}(E,E)+\sum_{\alpha=3}^4  {\cal M}^{\alpha,\alpha}(E,E)=0.
\end{equation}

\section{Details on the calculation of the noise} \label{apnoise}
The noise correlation function at a voltage $V$ is defined as
\begin{eqnarray}
S(eV)&=&\int_{-\infty}^{\infty} d\tau S(t,t-\tau), \nonumber \\
S(t,t^{\prime})&=&\langle \left[\delta J(t) \delta J(t^{\prime})+\delta J(t^{\prime}) \delta J(t) \right]\rangle, 
\end{eqnarray}
with
$\delta J(t) = \dot{N}_{\rm N}(t) - J(t)$.

Following a similar procedure as with the current, we evaluate
\begin{eqnarray}\label{S}
S &=&\frac{e^2}{4 h^2} \sum_{\alpha_1,\alpha_2,\beta_1,\beta_2} \int dE_1 dE_2 dE_3 {\cal M}^{\alpha_1 \alpha_2}(E_1, E_2) {\cal M}^{\beta_1 \beta_2} (E_3,E_3)\nonumber \\
& & \times \langle \left[\Psi^{\alpha_1}_{i}(E_1)\right]^{\dagger}  \Psi^{\alpha_2}_{i}(E_2)  \left[\Psi^{\beta_1}_{i}(E_3)\right]^{\dagger}  \Psi^{\beta_2}_{i}(E_3) \rangle - J J.
\end{eqnarray}
We analize
\begin{eqnarray}
& & \sum_{\alpha_1, \alpha_2,\beta_1,\beta_2} \langle \left[\Psi^{\alpha_1}_{i}(E_1)\right]^{\dagger}  \Psi^{\alpha_2}_{i}(E_2)  \left[\Psi^{\beta_1}_{i}(E_3)\right]^{\dagger}  \Psi^{\beta_2}_{i}(E_3) \rangle  \nonumber \\
& & = \sum_{\alpha_1, \alpha_2 \beta_1,\beta_2}
\langle \left[\Psi^{\alpha_1}_{i}(E_1)\right]^{\dagger}  \Psi^{\beta_2}_{i}(E_3)  \rangle \langle \Psi^{\alpha_2}_{i}(E_2)  \left[\Psi^{\beta_1}_{i}(E_3)\right]^{\dagger}  \rangle 
+ \ldots,
\end{eqnarray}
where $\ldots$ denotes a term that cancels out with $JJ$ in Eq. (\ref{S}). The other terms are
\begin{eqnarray}
& &   \langle \left[\Psi^{\alpha_1}_{i}(E_1)\right]^{\dagger}  \Psi^{\beta_2}_{i}(E_3)  \rangle \langle \Psi^{\alpha_2}_{i}(E_2)  \left[\Psi^{\beta_1}_{i}(E_3)\right]^{\dagger}  \rangle = \nonumber \\
& & \;\;  \;\;\delta(E_1-E_3) \delta(E_2-E_3) 
\delta_{\alpha_1,\beta_2}
\delta_{\alpha_2,\beta_1} f_{\alpha_1}(E_1)\left[ 1- f_{\alpha_2}(E_2) \right].
\end{eqnarray}

\section{Details of the calculation of the transfer matrix}\label{det-her}
 Following the procedure explained in Ref. \cite{van2011spin}, we  define the operator
\begin{equation}\label{jt}
\tilde{\rm v}{\cal J}=\frac{\partial H}{\partial p} \equiv {\rm v}  \tau^0\otimes \sigma^0  + {\rm v}_{\Delta}  \tau^x \otimes \sigma^y,
\end{equation}
which transforms the original Hamiltonian to an Hermitian one,
\begin{eqnarray}
& & \tilde{H}_{BdG}(x)={\cal J}^{-1/2} H_{BdG}(x) {\cal J}^{-1/2} 
\end{eqnarray}\label{mphb}

Given the operator defined in Eq. (\ref{jt}), we can calculate
\begin{equation}\label{j12b}
{\cal J}^{-1/2} = a \tau^0\otimes \sigma^0+b\tau^x \otimes \sigma^y,
\end{equation}
with the result
\begin{equation}\label{coefj12}
a=\pm\frac{1}{\sqrt{2 \tilde{\rm v}} }\sqrt{{\rm v}\pm\tilde{\rm v}},\;\;\;\;b=\mp\frac{{\rm v}_\Delta}{\sqrt{2 \tilde{\rm v}} \sqrt{{\rm v}\pm\tilde{\rm v}}},\;\;\; \tilde{\rm v}=\sqrt{{\rm v}^2-{\rm v}_{\Delta}^2}.
\end{equation}
Therefore,
\begin{eqnarray}
 & & \tilde{H}_{BdG}(x)=- i \partial_x \tilde{\rm v} \tau^0\otimes \sigma^0  - \varepsilon_0 \left(a^2 -b^2\right) \tau^z\otimes \sigma^0 \\
& & +\Delta_0\left[\left(a^2+b^2\right) \tau^x\otimes \sigma^0 +2 ab \tau^0\otimes \sigma^y\right]\nonumber \\
&=&- i \partial_x \tilde{\rm v}  \sigma^0 \otimes \tau^0  - \varepsilon_0 \tau^z\otimes \sigma^0
+\Delta_0\left[\frac{\rm v}{\tilde{\rm v}} \tau^x\otimes \sigma^0 - \frac{{\rm v}_{\Delta}}{\tilde{\rm v}} \tau^0\otimes \sigma^y\right].
\nonumber 
\\
\end{eqnarray}\label{mph1b}

The transfer matrix is calculated from
\begin{eqnarray}
& & \tilde{H}_{BdG}(x) \tilde{\Psi}(x)=E {\cal J}^{-1}  \tilde{\Psi}(x),\nonumber \\
& & {\cal J}^{-1} =\frac{1}{\tilde{\rm v}}
\left( {\rm v}  \tau^0\otimes \sigma^0  - {\rm v}_{\Delta}  \tau^x \otimes \sigma^y\right),
\end{eqnarray}
and $ \tilde{\Psi}(x)= {\cal J}^{1/2} \Psi(x)$,
with $\tilde{\rm v}=\sqrt{{\rm v}^2-{\rm v}_{\Delta}^2}$, where we focus on ${\rm v}>{\rm v}_{\Delta}$. 
 Hence $ \tilde{\Psi}(x_1)= M(E)\tilde{\Psi}(x_2) $. The result is Eq. (11).

\bibliography{super-hall.bib}

\end{document}